\documentclass{aa}  
\usepackage{graphicx}
\usepackage{capt-of}
\usepackage{color}
\usepackage{pdflscape}
\usepackage[utf8]{inputenc}
\usepackage[T1]{fontenc}
\usepackage{soul}
\usepackage{lastpage}
\usepackage{txfonts}
\usepackage{hyperref}
\usepackage{lscape}
\definecolor{brightpink}{rgb}{1.0, 0.0, 0.5}

\definecolor{electricblue}{rgb}{0.0,0.5,1.0}

\newcommand{\isotope}[2]{$^{#1}${#2}}

\newcommand{\msun}{$\rm{M_{\odot}}$}

\begin{document} 

   \title{Modeling the progenitors of low-mass post-accretion binaries}

    \author{A. Dimoff \inst{1}\fnmsep\inst{2}
          \and
          R.~J. Stancliffe \inst{3}
          \and 
          C.J. Hansen \inst{1}
          \and
          R. M. Seeburger \inst{2}
          \and
          H. Taylor \inst{3}
          }
    \institute{Goethe Universit\"at Frankfurt Institut f\"ur Angewandte Physik,
            Max-von-Laue-Str. 1, 60438 Frankfurt am Main Germany
        \and
            Max Planck Institute f\"ur Astronomie,
            K\"onigstuhl 17, 69117 Heidelberg Germany
        \and
            H.H. Wills Physics Laboratory, University of Bristol, 
            Tyndall Avenue, Bristol BS8 1TL, UK \\
        \email{dimoff@mpia.de}
            }
    

  \abstract
   {
   About half of the mass of all heavy elements with mass number A $>$ 90 is formed through the slow neutron capture process (s-process), occurring in evolved asymptotic giant branch (AGB) stars with masses $\sim1-6$ \msun. The s-process can be studied by modeling the accretion of material from AGB stars onto binary barium (Ba), CH, and carbon-enhanced metal-poor (CEMP)-s stars.
   }
   {
   Comparing observationally derived surface parameters and 1D-LTE abundance patterns of s-process elements to theoretical binary accretion models, we aim to understand the formation of post-accretion systems. We explore the extent of dilution of the accreted material and describe the impact of convective mixing on the observed surface abundances.
   }
   {
   We compute a new grid of 2700 accretion models for low-mass post-accretion systems. A maximum-likelihood comparison determines the best fit models for observational samples of Ba, CH, and CEMP-s stars. 
   }
   {
   We find consistent AGB donor masses in the mass range of $2 - 3$\msun$\,$across our sample of post-accretion binaries. We find the formation scenario for weak Ba stars is an AGB star transferring a moderate amount of mass ($\leq 0.5$ \msun) resulting in a $\sim 2.0-2.5$ \msun$\,$Ba star. The strong Ba stars are best fit with lower final masses $\sim1.0-2.0$ \msun $\,$and significant accreted mass ($\geq 0.5$ \msun). The CH and CEMP-s stars display lower final masses ($\sim1.0$ \msun) and small amounts of transferred material ($\sim 0.1$ \msun). 
   }
   {
   We find that Ba stars generally accrete more material than CEMP-s and CH stars. We also find that strong Ba stars must accrete more than $0.50$ \msun $\,$to explain their abundance patterns, and in this limit we are unable to reproduce the observed mass distribution of strong Ba stars. The mass distributions of the weak Ba stars, CEMP-s, and CH stars are well reproduced in our modeling. 
   }
   

   \maketitle

\section{Introduction}\label{sec:INTRO}

Barium (Ba) stars, CH stars, and carbon-enhanced metal-poor (CEMP)-s stars are recognized as post-accretion binary systems showing enrichment in carbon and elements synthesized mostly through the s-process \citep[][]{1942ApJ....96..101K, Bidelman_Keenan_1951_BaIIStars, 1957RvMP...29..547B, Kappeler+2011_review, Lugaro_etal2023}. Initially described as first ascent giants, the Ba, CH, and CEMP-s stars are not able to produce s-process material and self-enrich their envelopes. Additionally, dwarf stars have been observed carrying these signatures. 

Discovered in the 1950s \citep{Bidelman_Keenan_1951_BaIIStars, Bidelman1957_BaStars}, Ba stars show enhancements in s-process elements. Detailed chemical abundance analyses have confirmed the s-rich nature of the Ba stars \citep{Warner19656_BaStars, 2006A&A...454..895A, 2011A&A...533A..51P, DeCastro_etal2016, Cseh_Ba_etal2018, 2018A&A...618A..32K, Roriz2021, Roriz_etal2021}. The Ba stars characteristics include moderate to high metal content ([Fe/H] $>$ -1.0), s-element enrichment, and mild carbon enrichment. The binary nature of these systems was recognized by \citet{McClure+1980_Ba_binaries} and \citet{McClure1984}, and \citet{1990ApJ...352..709M} and \citet{1998A&AS..131...25U} later confirmed the link between the Ba stars and binary interactions with refined orbital parameters. More recent studies \citep{2019A&A...626A.128E, 2019A&A...626A.127J} have contributed to the compilation of orbital properties and component parameters for increasingly large samples of Ba stars. 

\citet{Keenan1942_CHstars} distinguished the CH stars from ordinary carbon stars by the strong CH molecular G-band near 4308 \AA, and \citet{Bidelman_Keenan_1951_BaIIStars} provided further classification based on their heavy element enrichment. Photometric surveys expanded the known population of CH stars, confirming their primary existence in the Galactic halo \citep{Bond1974_CHstars}. The CH stars share many properties with the Ba stars, but are found with higher carbon enhancements above [C/Fe] $>$ +0.5 and may be found at lower metallicities (-2.50 $<$ [Fe/H] $<$ -0.20). Recent studies \citep{Goswami_etal2006, Karinkuzi_Goswami2014, Karinkuzhi_Goswami2015, Goswami_Aoki_Karinkuzhi2016, 2019MNRAS.486.3266P}, expand on the heavy element abundance patterns. \citet{McClure1984} showed that CH stars exhibit long-term radial velocity variations, and show a very high binary fraction in the population. \citet{1990ApJ...352..709M} established that the CH stars have orbital properties consistent with mass transfer from an evolved asymptotic giant branch (AGB) companion with typically nearly circular orbits. The CH stars are a key population of stars, bridging the gap between the metal-rich Ba stars and the metal-poor CEMP-s stars.

The subclass of CEMP-s stars, identified by their enhanced surface abundances of s-process elements, were distinguished as a unique group among CEMP stars \citep{Ryan+1996_CEMPstars, Norris+1997_CEMPs}. The CEMP-s stars reside at low metallicities ([Fe/H] $<$ -1.0, but often much lower, $<$ -2.0), have a high carbon-to-iron ratio ([C/Fe] $>$ +0.7), and have strong over-abundances of s-process elements like barium ([Ba/Fe] $>$ +1.0). One of the most significant discoveries regarding CEMP-s stars is that nearly all of them are in binary systems, with a binary fraction near 100\%, providing strong evidence of mass-transfer as the source of the chemical enrichment \citep{2005ApJ...625..825L,Starkenburg+2014_CEMP_binaries,HansenT+_CEMP_binaries}. The binary nature of CEMP-s stars is widely supported by data acquired from extensive programs of radial velocity monitoring where orbital periods range from hundreds to tens of thousands of days \citep{1998A&A...332..877J, HansenT+_CEMP_binaries}. With refined orbital parameters of the observed population, \citet{Izzard+2010}, \citet{Starkenburg+2014_CEMP_binaries}, \citet{Hansen+2016_CEMPabunds_formation}, \citet{HansenT+_CEMP_binaries}, and \citet{Abate+2018_CEMP_binaries} show consistency with binary mass-transfer models. By studying the orbital properties of both CH and CEMP stars, \citet{Jorissen+2016_BinaryCH_CEMP} found no discernible differences in the period-eccentricity distribution of the two groups. They remark that the two classes of stars should not be treated separately from the orbital perspective, but as one population. As similar post-mass-transfer systems at different metallicities, these three populations of stars provide valuable observational constraints on s-process nucleosynthesis models in AGB stars \citep{Hansen+2016_CEMPabunds_formation, HansenT+_CEMP_binaries, Cseh_Ba_etal2018, Hansen_SrBa_etal2019, 2022A&A...660A.128C}. 

The peculiar abundance patterns in these populations are understood to be the result of AGB mass-transfer in binary systems. To explain the origins of the chemical enhancements in these stars, various models of accretion and mass transfer have been developed with the aim of determining how AGB stars enrich their binary companions, the efficiency of accretion, and evolutionary links between classes of chemically peculiar stars. \citet{1988A&A...205..155B} showed that for some Ba stars with short orbits, Roche-lobe overflow (RLOF) is a likely mechanism of mass transfer. However, many observed Ba, CH, and CEMP-s stars have long orbital periods where RLOF is not likely to occur. 

In the Bondi-Hoyle-Lyttleton (BHL) accretion regime \citep[][]{1944MNRAS.104..273B, 2002MNRAS.329..897H, 2004NewAR..48..843E} mass is captured from a stellar wind, where early models predicted inefficient accretion. Modern hydrodynamical simulations \citep{Mohamed_Podsiadlowski2007} show that wind Roche-lobe overflow (WRLOF) can significantly increase accretion efficiency, and can explain the enrichment observed in Ba, CH, and CEMP-s stars, even in wide binaries. Some models suggest accreted material can form a circumstellar disk around the secondary, allowing retention of enriched material. \citet{1995MNRAS.277.1443H} investigated the formation of Ba and CH stars via binary interactions, considering wind accretion, stable RLOF, and common-envelope evolution and ejection. Other modeling efforts in binary mass loss and accretion have been successful in generalizing the effects of interacting binaries \citep{Karakas+2000_Ba_ecc}.

After material has been accreted, mixing processes within the accretor star will affect the observed surface abundances. Convective mixing becomes more important with the advance of the convective envelope as the star ascends the giant branch. \citet{2011ApJS..197...17C} showed that the third dredge-up is more efficient at lower metallicities, making overall s-process enrichment weaker in the metal-rich Ba stars. Other mixing processes are explored by \citet{Matrozis_Stancliffe2016}, who showed how thermohaline mixing and dilution affect observed abundances in CEMP-s stars after accretion. 

The nucleosynthetic s-process is expected to take place in the interiors of low- to intermediate-mass (1-6 \msun) thermally-pulsing AGB (TPAGB) stars \citep[][]{1998ApJ...497..388G, Busso_Gallino_Wasserburg99, 2006NuPhA.777..311S, 2014PASA...31...30K}. In the framework of post-mass-transfer binaries, the more massive primary TPAGB star loses mass and pollutes the atmosphere of the less evolved companion \citep{2024AJ....167..184R}. At very low metallicities ([Fe/H] $\lesssim$ -2.5), there are insufficient seed nuclei for the s-process to operate, and s-process enhancement from AGB stars is not expected at such metallicities \citep{2014A&A...568A..47H, Lombardo+2025}. \citet{2014PASA...31...30K} modeled yields of s-process elements in AGB stars and showed that low-mass (1-3 \msun) AGB stars are primarily responsible for the chemical signatures observed in these post-mass-transfer systems. The FUll-Network Repository of Updated Isotopic Tables \& Yields (FRUITY) models \citep[][]{2006NuPhA.777..311S,2011ASPC..445...57D,2015ApJS..219...40C} provide updated s-process element yields from AGB stars at different masses and metallicities, making them useful for predicting s-process abundance patterns in Ba, CH, and CEMP-s stars, determining neutron source efficiency, and comparing observed stellar abundances to theoretical yields. 

The mass distribution of Ba stars was studied by \citet{2017A&A...608A.100E} and is described by two Gaussians, with a main peak at 2.5 \msun$\,$with a standard deviation of 0.18 \msun, and a broader tail at higher masses (up to 4.5 \msun), which peaks at 3 \msun$\,$with a standard deviation of 1 \msun. A complementary study in \citet{2019A&A...626A.127J} found a similar distribution of masses with compatible metallicities for Ba stars. Post-accretion binaries play a crucial role in stellar population synthesis and binary population synthesis modeling. 

The observable surface abundances of these stars are strongly dependent on the amount of mass transferred $\Delta M$, when the mass is transferred during the evolution of the secondary, and the mixing processes within the accretor star. One of the outstanding questions is how much material is actually accreted by the companion, which provides a direct link to the efficiencies of AGB mass transfer. \citet{Stancliffe2021} has modeled Ba star progenitor systems at a fixed metallicity of $\rm{[Fe/H]} = -0.25$ and a fixed post-accretion Ba star mass of $2.5\;\rm{M_{\odot}}$, corresponding to the average mass of Ba giants determined by \citet{2017A&A...608A.100E}. Under these assumptions, \citet{Stancliffe2021} found that the amount of accretion in Ba star systems is typically small, on the order of $0.10$ \msun. 

Our grid of stellar evolution models spans a wider range in mass and metallicity, compatible with both metal-poor CEMP and CH stars, and metal-rich Ba stars. Such models give insight into the progenitor system including: the mass of the primary AGB star, how much material is transferred, and the initial mass of the observed polluted star. We compare observationally derived stellar surface parameters and abundances to our evolution models. When viewed together as multiple populations formed through similar scenarios, these systems trace the formation processes of heavy elements produced in AGB stars and accretion in low-mass binary systems. We comment on the orbital properties of these populations in the literature. 

\section{Observational data}

\begin{figure}[!h]
    \centering
    \includegraphics[width=0.95\linewidth]{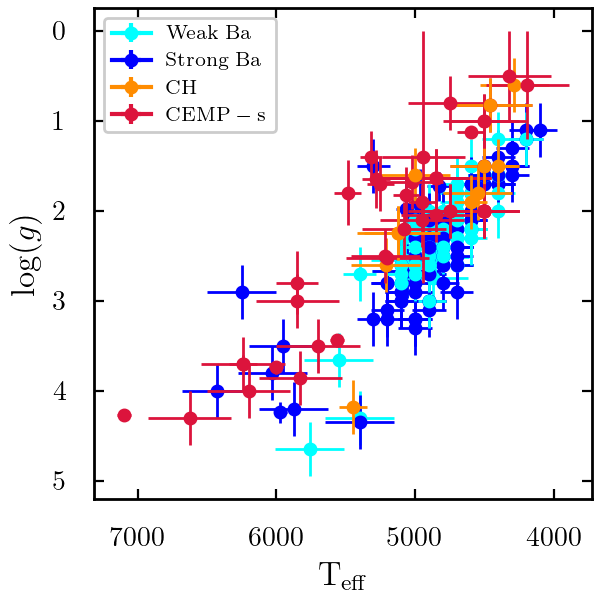}
    \caption{Kiel diagram showing surface gravities and effective temperatures for the collected observational sample. Blue data points are strong Ba stars, cyan data points are weak Ba stars, orange data points are CH stars, and red data points are CEMP-s stars. Surface parameters and abundances are collected from \citet{Dimoff+2024}, \citet{DeCastro_etal2016}, \citet{Roriz2021}, \cite{Goswami_etal2006}, \citet{Karinkuzi_Goswami2014}, \citet{Karinkuzhi_Goswami2015}, \citet{Goswami_Aoki_Karinkuzhi2016}, \citet{Goswami+2021_CEMPs}.}
    \label{fig:obs_kiel}
\end{figure}

We compare our models to stellar parameters and surface abundances derived from high-resolution (30000 $<$ R $<$ 60000) spectroscopic observations. We include the homogeneously analyzed sample of Ba, CH, and CEMP-s stars from \citet{Dimoff+2024}. We collect further data from the large sample of strong- and weak Ba stars from \citet{DeCastro_etal2016}, supplemented by additional heavy element abundances from \citet{Roriz_etal2021}. We expand our analysis to the metal-poor regime and include CH and CEMP-s stars from \cite{Goswami_etal2006, Karinkuzi_Goswami2014, Karinkuzhi_Goswami2015, Goswami_Aoki_Karinkuzhi2016, Goswami+2021_CEMPs}.  For each of our observational samples, we collect stellar surface parameters $\rm{T_{eff}}$, $\log{g}$, and [Fe/H] (see Figure \ref{fig:obs_kiel}) and surface abundances of C and heavy elements produced via the s-process, as well as the r-process tracer Eu. Not all of the observational datasets contain the same derived elemental abundances. In some cases, the observable Ba lines are saturated and not used for abundance derivations. By comparing our models to different populations of stars believed to have accreted material from a previous AGB companion, we are able to make predictions about AGB masses, stellar masses, and accretion masses, and make distinctions between metal-poor and metal-rich populations. These observational data are non-homogeneous and do not contain the same elements abundances, as we draw our observed data from different sources. 

We take note of stellar labels from the literature, and we apply corresponding labels based on metallicity and chemical enrichment. The original CEMP classification scheme from \cite{Beers_Christlieb_2005} and \cite{Masseron+2010_CEMP} defines metal poor stars as those with $\rm{[Fe/H]} < -1.00$. Within this category, we define CEMP stars in our sample as stars with $\rm{[C/Fe]} > 0.70$. For further classifying the stars in our sample, we define [ls/Fe] as the average abundances of Sr, Y, Zr, and Mo with respect to Fe, and [hs/Fe] as the average abundances of Ba, La, Ce, Pr, Nd with respect to Fe. From \citet{2014ApJ...787...10B} Nb is mostly produced by the s-process, but Nb abundances in the \citet{DeCastro_etal2016} dataset are systematically higher than the models, and we exclude Nb from the computation of [ls/Fe]. 

Following \citet{Hansen_SrBa_etal2019}, we classify our CEMP population based on relative ratios of light- and heavy- s-process element abundances. We substitute [ls/Fe] for [Sr/Fe] and [hs/Fe] for [Ba/Fe] to account for observations where derivations of these elements are missing. If the CEMP star has enhancements in both light and heavy s-elements ([Sr/Fe] or [ls/Fe] $>$ 0.30 and [Ba/Fe] or [hs/Fe] $>$ 0.30) and does not have significant r-process enhancements ([Eu/Fe] $<$ 0.50), it is a CEMP-s star. Otherwise we exclude any -r/s enriched stars from our investigation. The CH stars also have high carbon enhancements $\rm{[C/Fe]} > 0.70$, and are expected to have enhancements in s-process material with [ls/Fe] $>$ 0.30 or [hs/Fe] $>$ 0.30, but may have metallicities up to [Fe/H] $<$ -0.20. If there are no enhancements in either r- or s- elements, the star is labeled as a CEMP-no star. From our accretion modeling, we omit the CEMP-r, -r/s, and -no stars, as they are not believed to have accreted pure s-process signatures from a former AGB companion.

Ba stars have higher metallicities ($\rm{[Fe/H]} > -1.00$) and are generally not as enhanced in carbon as the CEMP and CH populations. Ba stars are split into two categories: strong- and weak Ba stars. \citet{Karinkuzhi_2021} provides a classification of the Ba stars based on heavy element abundance ratios. We use the [ls/Fe] and [hs/Fe] ratios to satisfy the criterion that more than three elements in the light- and heavy- s-process peaks are included. A star is a `Ba-no' star if both $\rm{[hs/Fe]}\; \rm{and}\; \rm{[ls/Fe]} < 0.20$. A `mild' (or `weak') Ba star has either [hs/Fe] or [ls/Fe] in the range $0.20 < \rm{[hs/Fe]}\; or\; \rm{[ls/Fe]} < 0.80$, and a `strong' Ba star has either $\rm{[hs/Fe]}\; or\; \rm{[ls/Fe]} > 0.80$. Our observational sample is displayed in Figure \ref{fig:obs_kiel}, where the stars are colored following these classification schemes. 

Some of our observational datasets contain alpha elements, including Mg. We find observed Mg abundances consistently higher than the AGB yield predictions, and the inclusion of Mg and other alpha elements is a hindrance in finding the best fit model, decreasing the likelihood that any given model fits the observed data. Previous studies on Galactic chemical evolution have also observed problems where the observations of Mg are systematically higher than the model predictions \citet{Romano+2010, Prantzos+2018, Jost+2024}. By this justification, we omit Mg and other alpha elements in the fitting routine, and focus only on C and the s-process material that is produced in the companion AGB star.

When explicitly stated, or when the uncertainties are not provided in the source material, we treat abundance upper limits by taking the value of the abundance to be consistent with zero (0), and the observed 'upper limit' is taken as the 2-$\sigma$ uncertainty. In this sense, the true uncertainty in the abundance is one half of value of the upper limit. 


\section{Modeling Methods}\label{sec:MODELING}

We compute a grid of binary accretion models using the \texttt{STARS} stellar evolution code \citep{1971MNRAS.151..351E, 1972MNRAS.156..361E, 1995MNRAS.274..964P, 2009MNRAS.396.1699S}. The program solves the equations of stellar structure with the chemical evolution of seven energetically important species: \isotope{1}{H}, \isotope{3}{He}, \isotope{4}{He}, \isotope{12}{C}, \isotope{14}{N}, \isotope{16}{O}, \isotope{20}{Ne}. At the convergence of every time step in the evolution model, a network of 40 isotopes are computed \citep{Stancliffe2005}. Our models use 499 mesh-points, no additional mass-loss, a mixing length of $\alpha = 2.025$ following \citet{Stancliffe2021}, and a prescription for convective overshoot following \citet{Schroder_etal1997}, using $\delta_{ov} = 0.15$ from \citet{Stancliffe_etal2015_overshoot}. We define the primary star to be the AGB donor star, and the secondary star to be the accreting Ba, CH, or CEMP-s star.

In our models, we do not consider thermohaline mixing, and the main mixing processes occur due to convection as the accretor star ascends the giant branch. We note that thermohaline mixing has been shown to be an effective means of changing the surface composition of low-mass stars on the upper part of the giant branch \citep{Stancliffe2015}. While \citet{Stancliffe+2007_THM} found that thermohaline mixing is effective on the main sequence, a follow-up study \citep{Stancliffe+2008_MixingCEMPs} found that the effects of gravitational settling can inhibit thermohaline mixing and, in cases where a small quantity of material is accreted, can be suppressed almost completely. \citet{Stancliffe2021} showed that thermohaline mixing plays an overall less significant role in diluting surface abundances compared to the convective mixing upon first dredge-up (FDU) in metal-rich stars. 

The grid of stellar evolutionary tracks is parameterized by the metallicity, the AGB donor mass, the initial mass, and the amount of accreted material. Metallicities range from $Z = 0.0001$ ([Fe/H]=-2.15) to $Z = 0.010$ ([Fe/H]=-0.15), complementary to the set of barium star models from \citet{Stancliffe2021} who computed models for metallicity of [Fe/H] = -0.25. It is assumed that the AGB donor and companion are the same metallicity. Our models cover a post-accretion mass range to include low- and intermediate-mass stars with final masses of $0.80$ to $5.0$ \msun$\;$.  The initial masses of the models are determined by subtracting reasonable amounts of material to be accreted. For any given final mass, initial masses range from $M_f - 0.50\;M_{\odot}$ to $M_f - 0.05$ \msun$\;$, with accretion masses equal to 0.05, 0.10, 0.20, 0.30, 0.40, and 0.50 \msun$\;$. In total, we generate around 2700 stellar evolution models. Following \citet{Stancliffe2021}, each stellar model is evolved up to an age that is determined by the lifetime of the AGB companion, where material from the AGB is accreted until the final target mass is achieved. Material is deposited on the surface and, after the accretion phase has ended, the secondary star continues to evolve taking the new surface composition into account. Surface abundances are diluted following mixing processes in the accretor star.

The composition of the AGB ejecta is taken from the FRUITY database \citet{2011ASPC..445...57D, 2011ApJS..197...17C, 2015ApJS..219...40C}. The FRUITY database contains around 120 models that range in initial mass from 1.3 - 6 \msun$\;$and metallicities from Z = 0.0001 ([Fe/H]=-2.15) to 0.03 ([Fe/H]=+0.32). Modeling AGBs gives the properties of the ejecta from these stars. However, these ejecta, once accreted onto a companion, will become diluted via the action of mixing mechanisms like convection. We use a tracer element `arbitrarium' to define how the AGB material is mixed into the secondary star. If the material originates from the AGB star, the value of arbitrarium is set to one, and it is set to zero if it originates within the secondary. We quantify the mixing following $$X_i^{actual} = X_{arb} X_i^{acc} + (1 - X_{arb})X_i^{original},$$ where $X^{actual}$ is the surface abundance, $i$ is the timestep, $X_{arb}$ is the value of arbitrarium, $X^{acc}$ is the ejecta abundance, and $X^{orignial}$ is the elemental abundance in a star of solar metallicity, taken from \citet{2009ARA&A..47..481A}, scaled by the metallicity of the star. The effects of mixing due to convection become more prominent with lower surface gravities and the inward advance of the convective envelope as the star ascends the giant branch. 

The time of accretion changes with the mass and metallicity of the AGB donor star, with more massive and more metal rich AGBs donating their material earlier than lower mass and more metal-poor stars. To this end, we first run the \texttt{STARS} code for single stars with masses of 1.3, 1.5, 2.0, 2.5, 3.0, 4.0, and 5.0 $M_{\odot}$, compatible with the AGB masses in the FRUITY database. Evolution in the \texttt{STARS} code continues past the core helium burning phase and begins the ascent of the AGB, past the evolutionary states of the observed stars. 

\begin{figure*}[!h]
    \centering
    \includegraphics[width=0.95\linewidth]{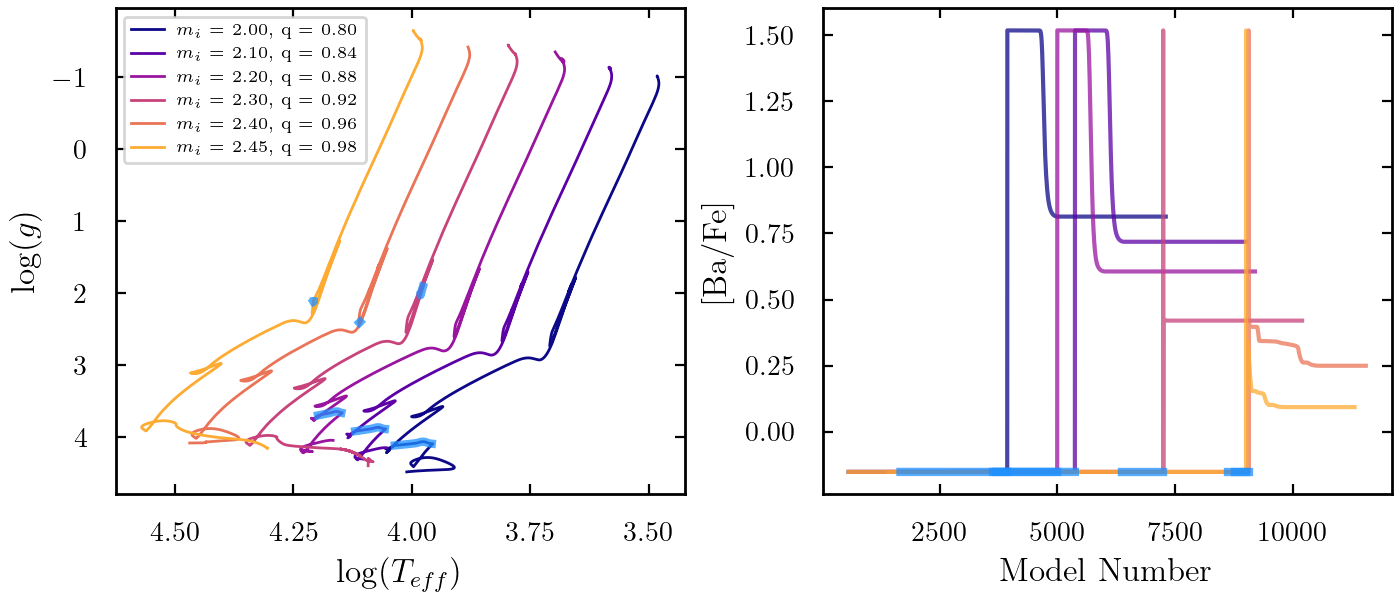}
    \caption{Left: Evolutionary tracks for a sample of stars with $m_{final} = 2.50$ \msun$\,$at [Fe/H] = -0.15 with different accretion masses and initial mass ratios, with accretion phases for each model highlighted in blue. Right: Relative surface abundance of the s-process element Ba. The abundance is elevated after the accretion phase, and only after the onset of first dredge-up and mixing is the surface abundance diluted.}
    \label{fig:acc_phase}
\end{figure*}

A selection of evolutionary tracks from our grid of models can be seen in Figure \ref{fig:acc_phase}, where in the left panel the evolutionary tracks are offset from one another in the Kiel diagram. The accretion phase is highlighted in blue. As a star gains mass, it will evolve more quickly, visible in the left panel of Figure \ref{fig:acc_phase}, where for the lowest initial mass tracks there is a noticeable change in temperature between the beginning and end of the accretion phase. The right panel of Figure \ref{fig:acc_phase} shows the surface abundance of [Ba/Fe], an element produced in AGB stars, as a function of model number, a proxy for time. Our models are parameterized such that after the star accretes the specified amount of mass, the surface abundances are changed to reflect that of the transferred AGB material. Beyond this, the convective envelope mixes and dilutes the surface abundances. 

\begin{figure}[!h]
    \centering
    \includegraphics[width=0.95\linewidth]{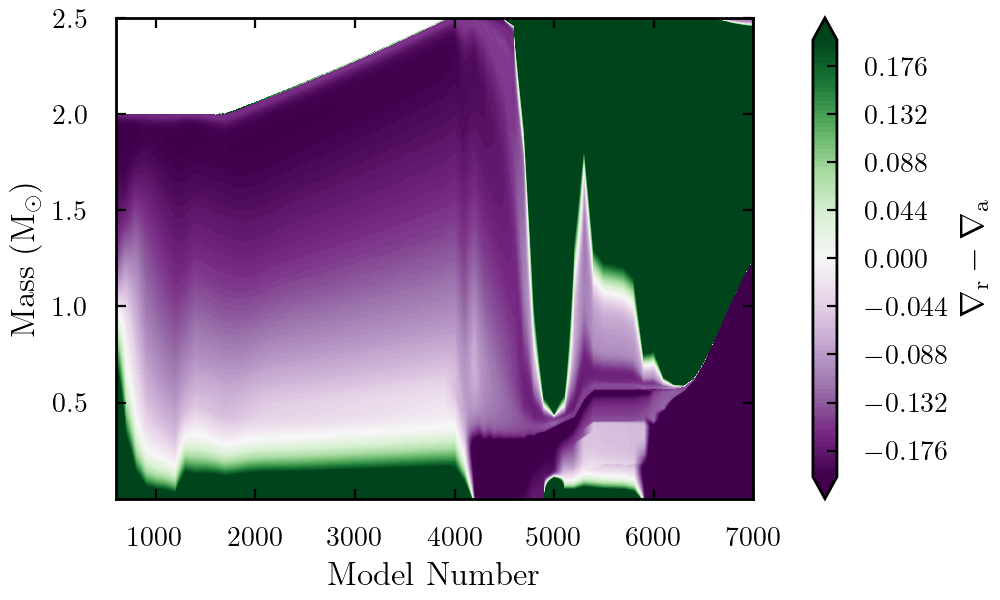}\\
    \caption{Example Kippenhahn diagram for a 2.00 \msun$\,$star accreting 0.50 \msun$\,$of material from a 2.50 \msun$\,$AGB star at a metallicity of [Fe/H] = -0.15. Green colors denote convective regions, and purple colors denote radiative regions, determined by the computed difference in the radiative and adiabatic transfer gradients. In the total mass on the y-axis, the accretion phase can be identified by the increase in mass. Note that the x-axis is model number, which is non-linear with respect to time.}
    \label{fig:Kippenhahn}
\end{figure}

An example Kippenhahn diagram is shown in Figure \ref{fig:Kippenhahn}. In this example, we display the structure of a 2.4 \msun$\,$star accreting 0.1 \msun$\,$of material from a 2.5 \msun$\,$AGB primary. The total mass is plotted against the model number, a proxy for time. The purple and green regions denote radiative and convective zones respectively, determined by computing the difference between the radiative and adiabatic transfer gradients, shown in the color bar. The onset of FDU is visible in the advance of the convective envelope towards the core as the star ascends the giant branch around model number 5000, as well as the onset of core He burning as the star begins to ascend the AGB after model number 5500. The accretion phase is shown as an increase in the overall mass of the star at the top of the plot. In this case, accretion occurs while the secondary star is on the main sequence.

We describe the general features of our grid of models by focusing on the case of a 2.00 \msun$\,$star accreting 0.5 \msun$\,$from a 2.50 \msun$\,$AGB star at a metallicity of [Fe/H] = -0.15 (seen in Figure \ref{fig:Kippenhahn}), which starts to transfer mass at a stellar age of around 700 Myr. At this point, the 2.00 \msun$\,$star reaches a core helium abundance of $\rm{X_{He}} \sim 0.45$, and the convective core contains around 0.50 \msun$\,$of material. As accretion progresses, the convective core expands to reach a mass coordinate around 0.65 \msun, where the core helium abundance is reduced as fresh hydrogen is ingested into the core of the now more-massive star. Since we are only accounting for mixing due to convection, the accreted layer remains on the stellar surface until the onset of FDU as the star ascends the giant branch. At its maximum depth, the convective envelope reaches a mass coordinate of around 0.50 \msun, and penetrates the outer layers of the core. This signifies the maximum dilution of the accreted material, as the envelope will not reach this depth for the remainder of the lifetime of the star. As in \citet{Stancliffe2021}, we can define the dilution factor $$d = M_{acc} / M_{mix},$$ where $M_{acc}$ is the mass of the accreted material and $M_{mix}$ is the mass through which it is mixed. At this point, $d = 0.25$. While expressed in different notation, this is analogous to the dilution factor described in \citet{2023A&A...672A.143D}.

As the initial mass ratio $q$ of the binary approaches one, the accretion happens later in the lifetime of the secondary star (see Figure \ref{fig:acc_phase}). A secondary star with a mass of 2.40 \msun$\,$accretes after it has left the main sequence and is going through the core-He burning phase. At this point, the convective envelope is considerably shallower than at the onset of FDU, and has a thickness of only about 0.50 \msun. Convection dilutes the accreted layer, and as the star begins to ascend the AGB the convective envelope again deepens, further diluting the accreted material. During this time, the maximum mass contained in the envelope is $\sim 1.90$ \msun. If the accretion occurs later on in the lifetime of the secondary star, mixing processes have already begun that significantly reduce the amount of accreted material remaining on the stellar surface.

The evolution of the surface abundances generally follows that of the dilution factor as temperatures in the envelope are not high enough to process the heavy elements accreted from the AGB star. In the right panel of Figure \ref{fig:acc_phase}, the surface abundances of the s-process element [Ba/Fe] are displayed. With only convection acting, the abundance of Ba is unchanged from the end of the accretion phase until the onset of FDU. With the deepening of the convective envelope, the accreted material is mixed and diluted within the stellar interior and the surface abundance falls dramatically. With the increase of the mass of the accretor star, the mass of the convective envelope increases, and through this the surface abundances are diluted more effectively. 

\subsection{Comparison to the models}

To quantify the comparison between our models and observations we compute the likelihood for each model at each time step, comparing the observed values of the stellar surface temperature, surface gravity, metallicity, and available abundances to those within the models. To find the best fit model, we maximize the likelihood function by minimizing the $\chi^2$ difference between the models and observations. One caveat to the $\chi^2$ method is that it generally assumes the parameters being fit are uncorrelated; in our comparison, this is not true. The relative abundances of heavy elements are highly correlated with one another and the surface parameters. In many cases, the difference in $\chi^2$ between best-fit models can be small, indicating only subtle differences between them. 

In this study, we consider models in the grid that have mass ratios $q = M_2 / M_1 \leq 1.00$. Otherwise, the companion will be more massive than the AGB star and will evolve first. We require that each observed star have measured surface parameters $\rm{T_{eff}}$, $\log(g)$, and $\rm{[Fe/H]}$, and computed 1D-LTE abundances for at least one s-process element in each s-process peak. Many evolutionary tracks exist on the HR diagram during periods of rapid evolution (i.e. the red giant branch (RGB), where many stars in our sample lie), and the density of model data points in these regions can be preventative in selecting a model that is physically representative of the observed system. Since this is the case, we exercise caution when addressing the best fit models using this method.


\section{Results}\label{sec:RESULTS}

\subsection{Fits to known Ba, CH, and CEMP-s stars}

We compare our model grid to the properties of known Ba, CH and CEMP-s stars. Comparing the effective temperature, surface gravity, metallicity, stellar mass, and elemental abundance patterns, we determine the most probable initial configuration to have produced the observations. We estimate the mass of the AGB star that transferred mass to the observed star, the initial and final mass of the chemically peculiar star, and the amount of accreted material required to reproduce the observed chemical signatures. We are able to recover the metallicity of the observed system with high accuracy. For each star, we visually examine the three best fit models according to their $\chi^2$ values. In many cases, there is only a small difference in $\chi^2$ between the best fit models. Before discussing the sample as a whole and the constituent populations of stars within our sample, we describe individual objects. \citet{Stancliffe2021} analyzed a subset of 74 Ba stars from \citet{DeCastro_etal2016}, although with fewer abundances compared to our inclusion of the data from \citet{Roriz2021}. Here we examine our own fits to some stars examined by \citet{Stancliffe2021}.

\begin{figure*}[!h]
    \centering
    \includegraphics[width=0.95\linewidth]{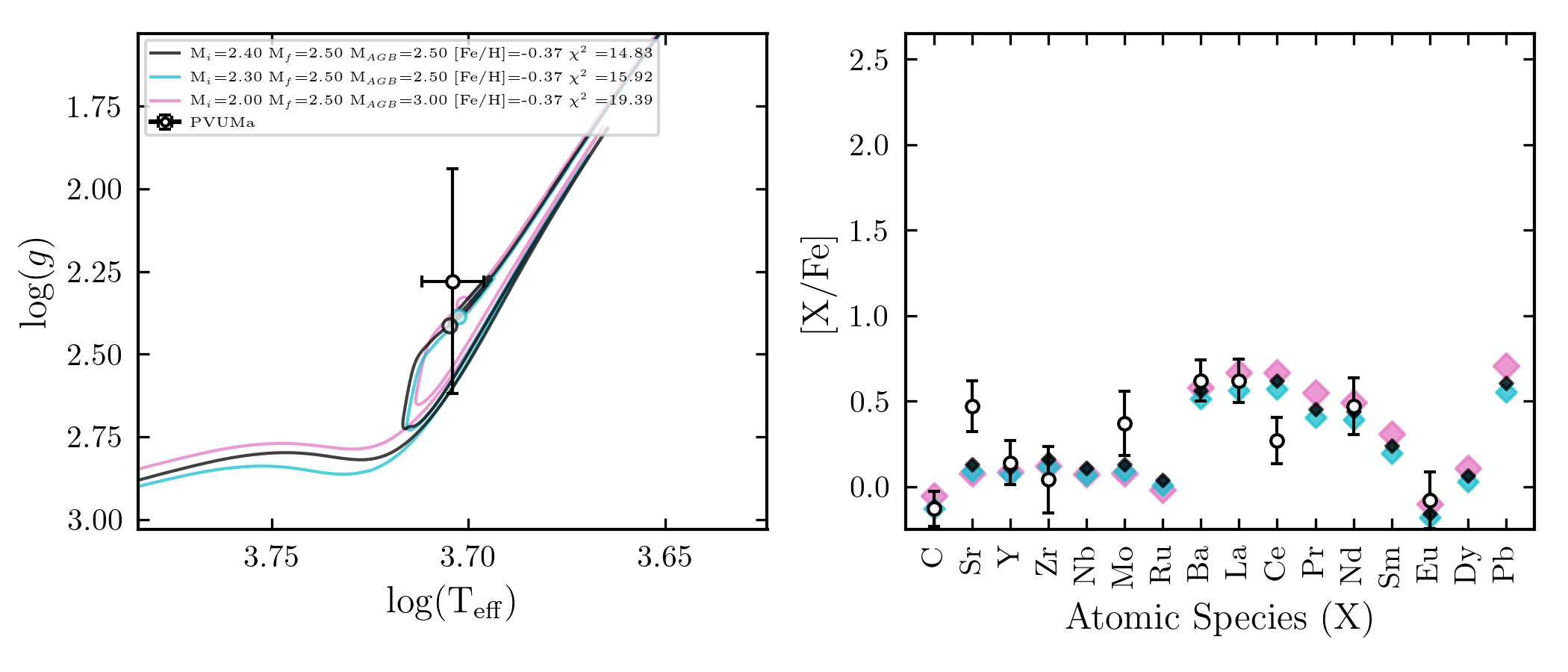} \\
    \includegraphics[width=0.95\linewidth]{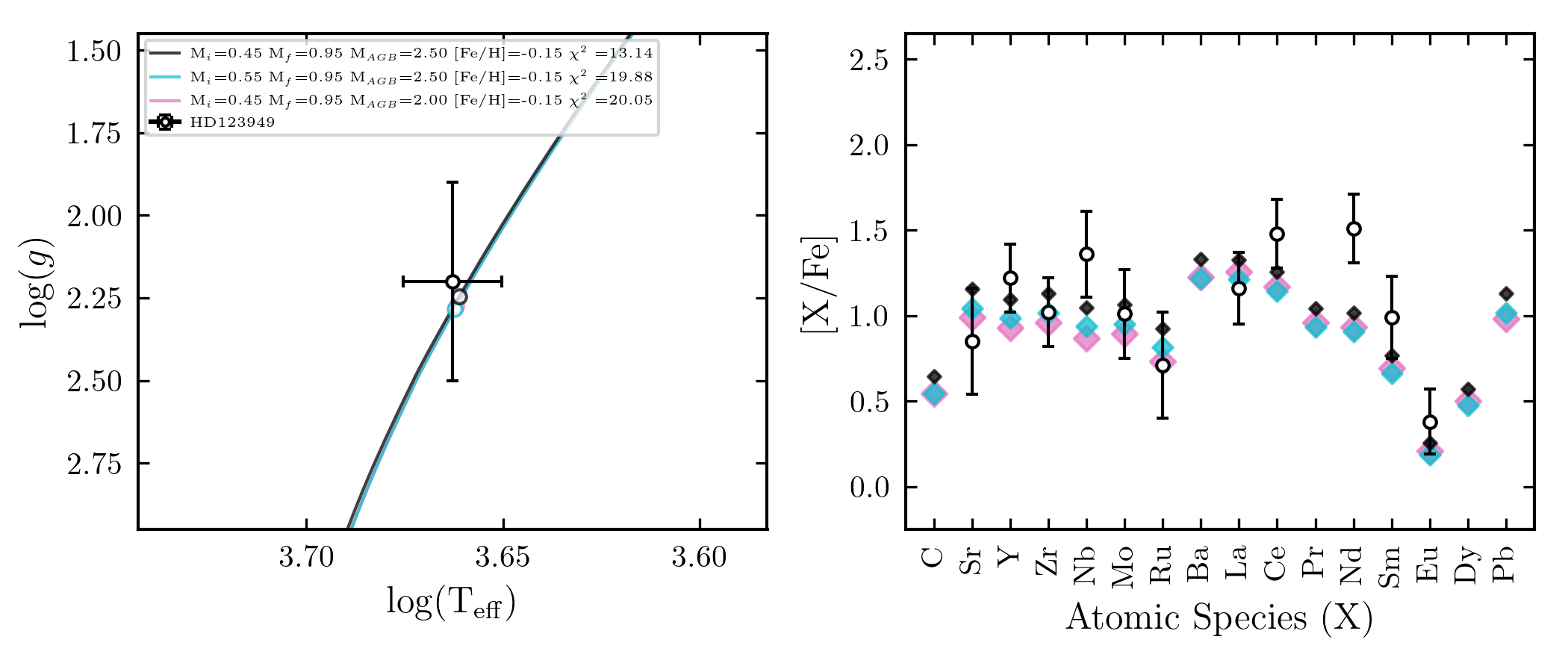} \\
    \includegraphics[width=0.95\linewidth]{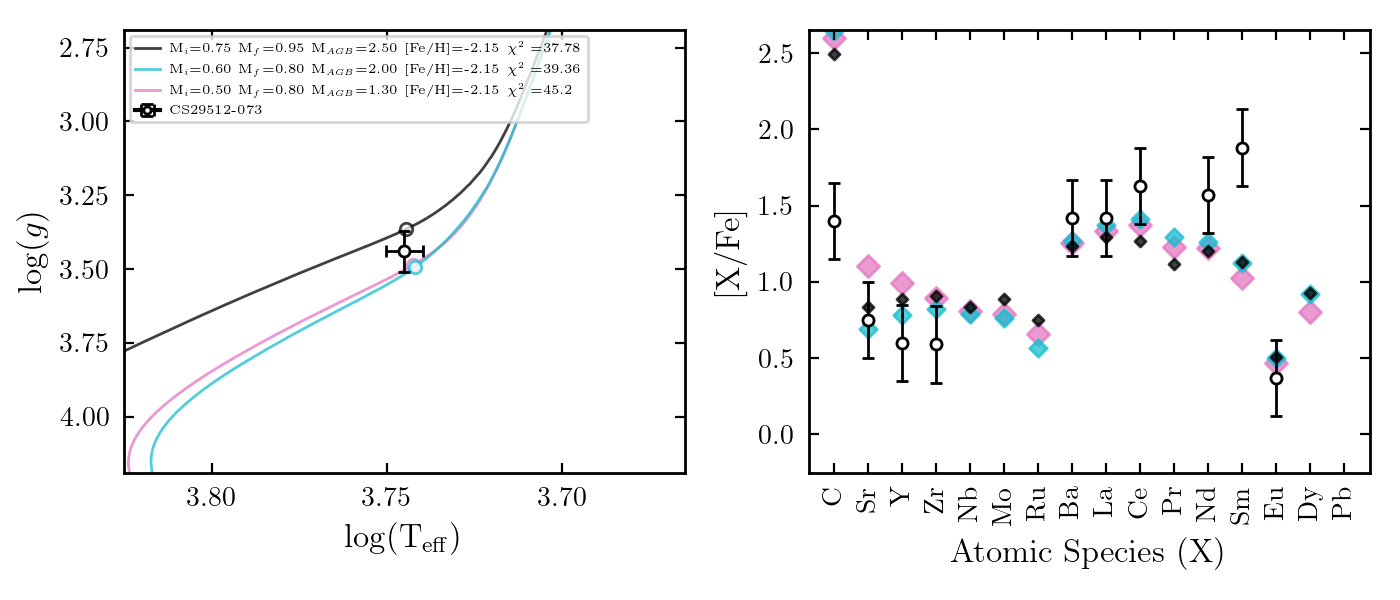} \\ 
    \caption{Kiel diagrams and abundance fits for the weak Ba star PV UMa, the strong Ba star HD 123949, and the CEMP-s star CS 29512-073, each showing the three best-fitting models and their associated $\chi^2$ values.}
    \label{fig:best_fits}
\end{figure*}

\paragraph{PV UMa} We find a very good fit for this weak Ba star in both the Kiel diagram and in abundance space, shown in the top panels of Figure \ref{fig:best_fits}. Our best fit to this star is a 2.40 \msun$\;$ star accreting 0.10 \msun$\;$ of material from a 2.50 \msun$\;$ AGB star at a metallicity of [Fe/H] = -0.37. The observed temperature The two best fit models could each be likely scenarios to produce this and other weak Ba star systems. Recently observed and studied by \citet{Dimoff+2024}, PV UMa is best fit on the second ascent of the giant branch, after the onset of core He burning. Computed abundances including carbon are well represented in the models, except for a over-abundance of Sr and an under-abundance of Ce. The low accretion mass and significant mixing on the giant branch results in relatively low surface abundances compared to the Ba strong stars

\paragraph{HD 123949} This object is representative of a large portion of the strong Ba stars in the \citet{DeCastro_etal2016} sample in both the Kiel diagram and in abundance space. An acceptable fit is obtained for this system, shown in the middle panels of Figure \ref{fig:best_fits}. This star lies directly on the evolutionary track for a 0.95 \msun$\,$star and the abundances are well-fit, save for slight over-abundances in Nb and Nd. The low $\chi^2$ values indicate a good fit, given the number of fitting parameters. We find best fit models to be low initial mass secondary stars $M_i \sim 0.45 - 0.55$ \msun, accreting significant ($> 0.40$ \msun) amounts of material from a 2.50 \msun$\,$AGB star resulting in a 0.95 \msun$\,$Ba giant. This does not fully agree with estimated Ba star masses from \citet{2017A&A...608A.100E}. More massive stars will have more massive convective envelopes, which will more efficiently mix and dilute surface abundances on the ascent of the giant branch, and a less massive accretor star will retain higher surface abundances. However, it may be an outlier in the strong Ba star mass distribution. 

\paragraph{CS 29512-073} In this CEMP-s star, we find an excellent match in metallicity with both observed and modeled [Fe/H] = -2.15. The precise observed stellar parameters are in good agreement with low-mass stellar evolution tracks on the sub-giant branch, with best fit final masses between $M_f = 0.80 \rm{to} 0.95$ \msun. These values are in good agreement with theoretical masses of metal-poor stars. Carbon is over produced by the models compared to the observations; thermohaline mixing on the main sequence could dilute the surface abundance of C. Abundances of heavy elements are generally fit by our modeling aside from a slight overabundance in Nd and an overabundance in Sm past the second s-process peak. The best fit AGB mass is 2.50 \msun, and our modeling suggests around $0.20$ \msun$\;$of material accreted from the former AGB companion. 
\\ 

\begin{figure*}[!h]
    \centering
    \includegraphics[width=0.95\linewidth]{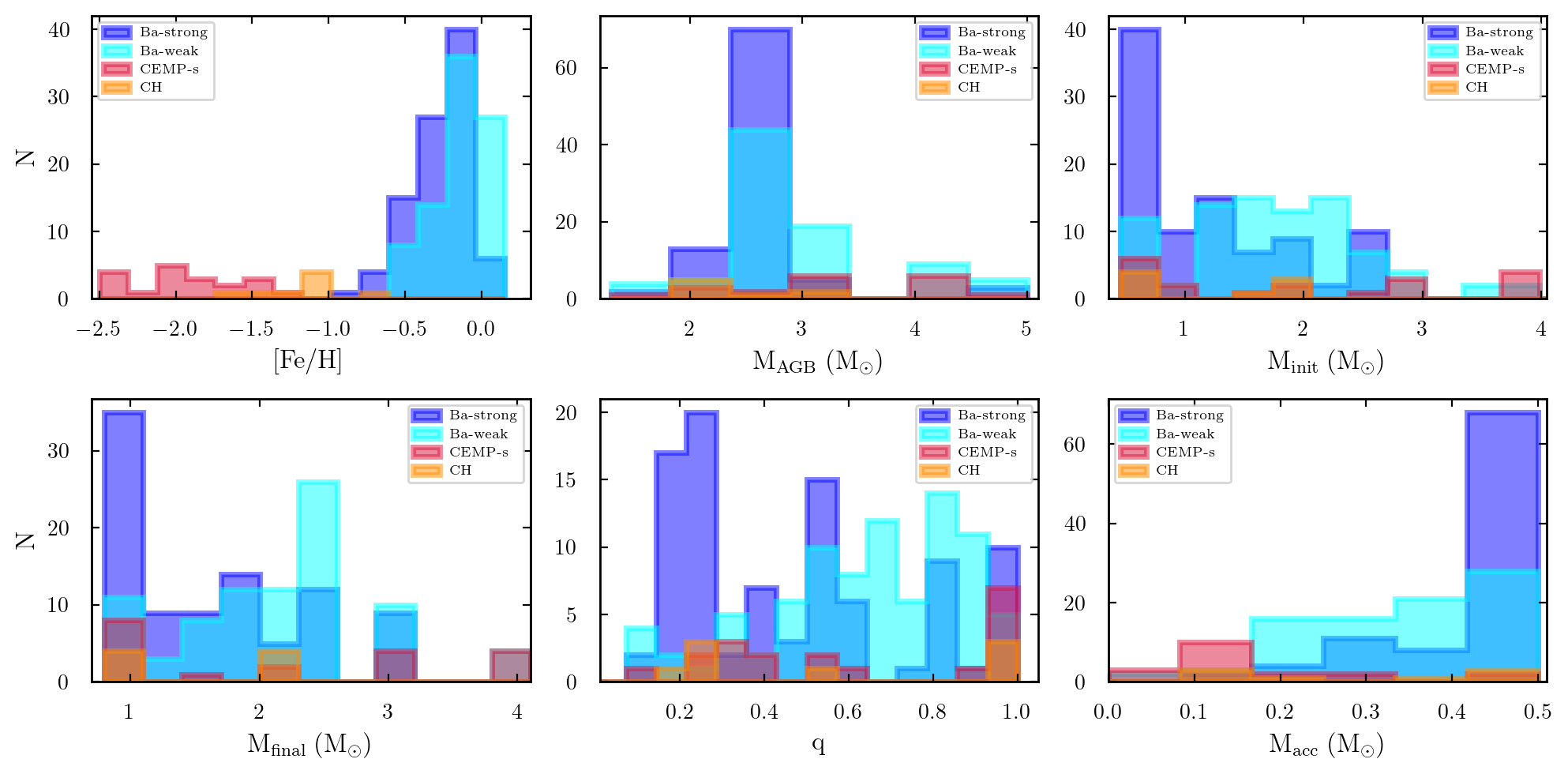}
    \caption{Histograms of model parameters for the different classes of stars in our investigation. Dark blue regions are strong Ba stars, cyan regions are weak Ba stars, red regions are CEMP-s stars, and orange regions are CH stars.}
    \label{fig:full_histogram}
\end{figure*}

We present histograms of model parameters for our complete sample in Figure \ref{fig:full_histogram}. The weak Ba stars are shown in cyan, the strong Ba stars in blue, the CEMP-s stars in red, and the CH stars in orange. While our grid is not continuously nor equally spaced in initial nor final mass, we are able to make generalized statements about the progenitors of these post-accretion systems. We recover the metallicity distribution of stars in our sample, and since our model grid is limited to metallicities above [Fe/H] = -2.15, we make a cut on the observed stars at [Fe/H] $>$ -3.00. 

Overall, there is a clear preference of donor AGB masses across all stellar classifications with a very strong peak with a mean of $\langle \rm{M_{AGB}}\rangle =$ 2.70 \msun. This is expected, as the s-process production rates peak in this mass range. Fits to the strong Ba stars show a very pronounced peak at an AGB mass of 2.50$\;$\msun, with a width of $\sigma 0.52$ \msun. For the weak Ba stars, most of the fits suggest an average AGB mass of $\langle \rm{M_{AGB}}\rangle =$ 2.80 \msun. The distribution is broader compared to the strong Ba stars, with a width of $\sigma = 0.78$ \msun. For both the carbon-enhanced CH and metal-poor CEMP-s populations, the preferred AGB mass is $\langle \rm{M_{AGB}}\rangle =$ 2.80 \msun$\;$with an even broader distribution ($\sigma = 0.88$). 

Across the four different populations of post-accretion systems analyzed, all display non-gaussian behavior in their initial and final mass distributions. The weak Ba stars show a broad distribution in initial mass $\langle M_{init} \rangle =$ 1.70 \msun$\;$ with a width of $\sigma_{M_{init}} = 0.74$, and final masses around $\langle M_{init} \rangle =$ 2.30 \msun, and $\sigma_{M_{final}} = 0.75$. The mass ratios $q$ of the weak Ba stars show a peak at higher values close to 0.90 with a tail down to low mass ratios of 0.10. The models suggest that weak Ba stars accrete moderate amounts of material; on average, they accrete about 0.35 \msun$\,$ from their AGB companion. 

In the strong Ba stars, we find a very dominant peak at low initial masses around 0.5 \msun$\,$with a tail extending up to 2.50 \msun. The final masses are concentrated at about 1.00 \msun$\,$with a tail extending as high as 3.00 \msun. The mass ratios $q$ of strong Ba stars is almost opposite that of the weak Ba stars, with most models fitting low $q$ values of 0.20. There is a long, weak tail extending to higher mass ratios, with a small peak at mass ratios close to $q =$ 1.0. The models suggest that for strong Ba stars, large amounts of material have been accreted, with most fits showing 0.50 \msun$\;$. The distribution is strongly peaked at high accretion masses, with few systems indicating smaller amounts of accretion. This is more material compared to the weak Ba stars, which is reflected in the observed higher surface abundances in the strong Ba stars. 

At intermediate metallicities, the CH stars show initial masses with peaks at $\approx$ 0.50 \msun$\;$and a small broader peak at $\approx$ 2.00 \msun. Final masses show corresponding peaks at $\approx$ 0.90 \msun$\;$and around 2.50 \msun. Mass ratios in the CH stars are distributed between low ($\approx 0.20$) and high ($\approx 0.98$) values. The distribution of accretion masses in the CH systems display no clear preference in accretion masses. The number of CH stars in our sample is likely too low to make any strong claims about the population as a whole. 

At the lowest metallicities, the CEMP-s stars are fit with bimodal distribution in initial and final masses. The initial masses show peaks at $\approx$ 0.50 \msun$\;$and $\approx$ 2.0 \msun$\;$with a tail extending to higher initial masses. The final masses show a peak around 0.90 \msun$\;$, with a second peak close to 2.70 \msun. As in the initial masses, there is a tail extending to higher masses. This could be for the dwarf stars, which are not as well modeled as the giant stars. Initial mass ratios for the metal-poor systems show two clear peaks at $q \approx 0.30$ and $q \approx 1.00$, with a preference for higher initial mass ratios. Our modeling finds low accretion masses around 0.05 \msun for the CEMP-s stars.

\section{Discussion}\label{sec:DISCUSSION}

The strong Ba stars almost always accrete the maximum amount of mass, and the final mass distribution we recover does not fully agree with that of \citet{2017A&A...608A.100E}, where we find lower initial and final masses for many strong Ba stars by about a full solar mass. At higher masses, the more massive convective envelope dilutes surface abundances more efficiently compared to lower masses, and to maintain the observed higher surface abundances in the strong Ba stars, lower masses are required. We note our sample of Ba stars is only from \citet{DeCastro_etal2016} and \citet{Dimoff+2024}, where the larger \citet{2017A&A...608A.100E} sample includes the \citet{DeCastro_etal2016} data set, plus more Ba stars compiled from \citet{1983ApJS...52..169L,Lu_BaTaxonomy1991,1993A&A...275..101E,1996BaltA...5..217B}. The differences in average mass could be influenced by the larger sample. 

We find that when accretion happens while the secondary is on the first ascent of the giant branch, the mixing processes within the advancing convective region have a significant effect in quickly diluting the surface abundances. Our models generally suggest higher accretion masses in secondary stars accreting on the main sequence or while first ascending the giant branch, compared to those close to the tip of the RGB, or near the onset of core He burning. This is also due to evolutionary reasons, where in systems with mass ratios $q \approx 1$ the primary and secondary stars are of near equal mass and can evolve nearly simultaneously. 

Comparing our models and the observational data, we find a similar effect in the relative abundance uncertainties as the number of available abundances themselves. The average uncertainty in the \citet{Dimoff+2024} abundances is $\sigma_{[X/Fe]} \sim 0.15$, where in the combined \citet{DeCastro_etal2016} and \citet{Roriz2021} dataset it is $\sigma_{[X/Fe]} \sim 0.20$, and we find consistently smaller uncertainties approximately equally important when one has fewer computed abundances. The combined dataset from \citet{Cristallo+2016} (and references therein) has average abundance uncertainties of $\sigma_{[X/Fe]} \sim 0.23$, and the CEMP-s stars dataset from \citet{Goswami+2021_CEMPs} reports uncertainties of $\sigma_{[X/Fe]} \sim 0.25$, on average. 

In the evolution of low-mass metal-poor stars, we generally find the accretion happens while the secondary star is on the main sequence. With initial AGB masses much higher than accretor masses (2.50 \msun$\,$vs. $<$1.00 \msun), the secondary star is on the main sequence when the AGB star begins thermal pulsations. Without thermohaline mixing, the surface abundances remain nearly constant until the onset of first dredge-up, making the differentiation between best fit models in this regime difficult \citep{Stancliffe+2007_THM,Stancliffe+2008_MixingCEMPs}. Thus we find this set of models better for analyzing giant stars, and main sequence stars may not be as representative. 

We find it unlikely that all barium stars are observed shortly after the accretion process has ended, as would need be the case for initial mass ratios very close to $q = 1$, where mixing on the giant branch is very effective and quickly dilutes surface abundances (see the right panel in Figure \ref{fig:acc_phase}). Thus, larger accretion masses at lower initial mass ratios make sense for stars that exhibit large surface abundances of heavy elements. As the secondary star gains mass, it will alter its evolutionary path to follow that of the post-accretion mass. Some models in the grid have initial mass ratios $q > 1.0$ where the secondary star is more massive than and would evolve more quickly than the primary. While these models could provide useful in analyzing systems with multiple mass transfer events, we do not consider these models in this analysis. 

A few stars in our sample are poorly fit due to relatively low surface gravities ($\log(g) \leq 0.50$), very high surface abundances of heavy elements ([X/Fe] $>$ 2.00), large amounts of Eu compared to the s-process elements, or some combination of these. Very few evolutionary tracks with high elemental abundances extend toward the upper part of the AGB phase in the Kiel diagram, and the ones that do have high masses close to 5.00 \msun $\,$that do not agree with the theoretical masses of these stars. Fits resulting in high masses often also find flattened abundance patterns indicating the accreted material has been fully diluted, and the surface abundances have returned to the un-enhanced values. 

\subsection{Mixing and Dilution}

\citet{Matrozis_Stancliffe2016} found that atomic diffusion does not have a substantial effect on the surface abundances of CEMP-s stars, so the dilution of the accreted material, while variable in degree from one star to the next, is most likely the same for all elements, and the mass ratios and overall pattern will be preserved. Thermohaline mixing on the main sequence will dilute the surface abundances to some extent. \citet{Stancliffe2021} tested the inclusion and exclusion of thermohaline mixing in stars, and after the onset of FDU the surface abundance ratios of heavy elements between the two cases are nearly identical in Ba stars. Convective mixing processes in giant stars outweighs thermohaline mixing at higher metallicities, and the overall level of mixing in both scenarios is the same after FDU. In more metal-poor stars, this is not the case; \citet{Stancliffe+2007_THM} found that thermohaline mixing is on the main sequence is effective for light elements C, N, and O, and predicts different surfaces abundances for the CEMP-s stars compared to models with purely convective mixing. Including additional mixing processes, \citet{Stancliffe+2008_MixingCEMPs} found that the effects of gravitational settling can inhibit thermohaline mixing in CEMP-s stars and, in cases where a small quantity of material is accreted, mixing can be suppressed almost entirely. 

The accretion phase in our models coincides with the AGB phase of the donor, so the initial mass ratio of the system effectively controls when the accretion phase will happen during the lifetime of the secondary. If accretion occurs while the secondary is on the main sequence or on the sub-giant branch, the onset of first dredge-up (FDU) will severely deplete the surface abundances. If accretion happens after the onset of FDU, significant mixing will still occur in the convective envelope. \cite{2017A&A...606A.137M} suggests that in the wind regime only about 0.05 \msun$\,$of material can be added to the accreting star before it reaches critical rotation, which we find is able to explain the chemical enrichment of many CEMP-s stars, but not the Ba stars which require more mass to be accreted. According to \cite{2017A&A...606A.137M}, the specific angular momentum of the accreted material should be be lower than the Keplerian value by about an order of magnitude, or significant angular momentum losses must occur for substantial accretion to take place. This may not be a significant issue for the weak-Ba stars, where the larger radius of the giant accretor star requires more angular momentum transfer to reach the critical rotation velocity, allowing more mass to be accreted. For the strong-Ba stars, which tend to accrete significant amounts of material on the main sequence, this poses a problem, and BHL wind mass transfer is not applicable to these systems.

\subsection{Accretion efficiencies}

For our binary systems, we compute the accretion efficiency $\eta_{acc}$ by dividing the amount of mass gained by the accretor by the total mass lost by the AGB star in the system: $$\eta_{acc} = (m_{final} - m_{initial}) / (m_{AGB} - m_{WD}) \times 100\;[\%].$$ \noindent In this sense, we assume non-conservative mass transfer. We estimate white dwarf masses using tabulated H-exhausted core masses from \citet{2015ApJS..219...40C} for the given AGB mass and metallicity of the model. We compare the computed white dwarf masses to the dynamical estimates from \citet{Dimoff+2024} and find good agreement. We find no significant differences between our samples of CH and CEMP-s stars in this analysis of accretion efficiencies and in this instance we treat them as one metal-poor and carbon-rich population, in agreement with \citet{Jorissen+2016_BinaryCH_CEMP}

\begin{figure}[h]
    \centering
    \includegraphics[width=0.95\linewidth]{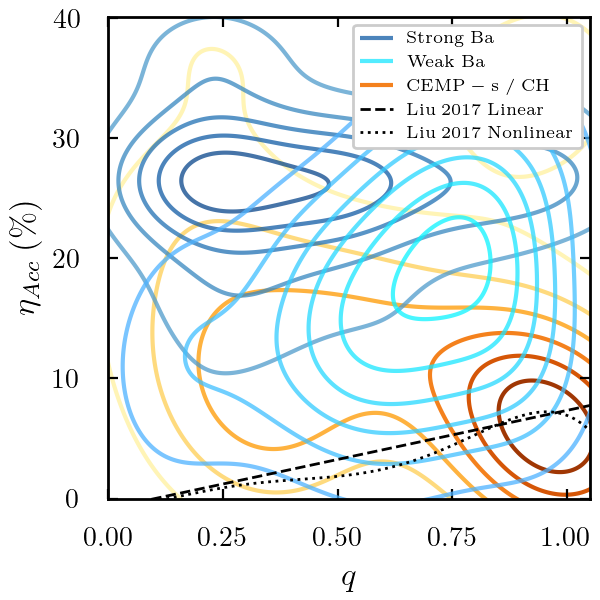}
    \caption{Computed accretion efficiencies for our sample populations. Purple-blue contours are the strong Ba stars, and the green contours are the weak Ba stars. Red-yellow contours are our carbon-enhanced sample, including both CH and CEMP-s stars.}
    \label{fig:acc_eff}
\end{figure}

We compare our computed $\eta_{acc}$ to 3D hydrodynamical wind mass-transfer models from \citet{2017ApJ...846..117L} in Figure \ref{fig:acc_eff}, where the black dashed and dotted lines are linear and 5th order polynomial fits from the 3D models respectively. These simulations generally result in low accretion efficiencies, with a trend toward higher efficiencies at higher mass ratios. 

The models in our grid are discrete points, and many times the same model is chosen to represent multiple modeled systems. For this reason, we present our populations as contours in $\eta - q$ space. We see a general trend of increasing efficiency $\eta$ with decreasing mass ratio $q$ across the different stellar populations. The metal-poor stars require less accretion in systems with initial mass ratios closer to $q = 1$ to reproduce the observed surface abundances, resulting in lower efficiencies. The bimodal distribution of the carbon-enhanced systems is well-shown in the density contours. At lower mass ratios below about $q \lesssim 0.5$, there are only a few carbon-enriched systems with relatively low mass-transfer efficiencies, $\approx 10\%$. Higher metallicity systems with lower mass ratios congregate in efficiencies around $\sim 25\%$; these are the strong Ba stars. Generally, we find that accretion masses greater than about $0.30$ \msun$\,$will result in efficiencies higher than the theoretical wind mass transfer regime described in \citet{2017ApJ...846..117L}.

\subsection{Possible mass transfer scenarios}

The mechanism of mass transfer in AGB stars is directly linked to the properties of the initial binary orbit and the relative velocity of the AGB wind compared to the orbital velocity. There are multiple regimes in which mass transfer can occur, and in all of them some amount of angular momentum transfer will perturb the orbit. The resulting orbit will likely not be the same as the initial orbit of the binary. 

\citet{2025arXiv250410939K} suggest that WRLOF and non-conservative mass-transfer with the presence of a circumbinary disc provides mechanisms for both high accretion efficiencies and enhanced eccentricities, explaining the observed orbital geometries of these post-accretion systems. For slow and dense stellar winds with low wind velocities compared to the orbital velocity, structures resembling wind Roche-lobe overflow form as the stars approach periastron, indicating a form of tidally enhanced mass transfer \citep[][]{Mohamed_Podsiadlowski2007,Mohamed_Podsiadlowski2012}. Through this process, the circumstellar outflows are non-spherical and concentrated in the orbital plane of the binary. The non-symmetric flows contribute to maintaining or increasing the eccentricity of the binary system. \citet{Abate+2013} discusses wind-RLOF in the context of binary population synthesis for the CEMP-s stars, finding that, pending the stability criterion, in short period systems RLOF is likely to occur. However, most $>70\%$ of the CEMP-s stars in their synthetic population have wide orbits and accrete through some form of wind-driven mass-transfer. We find that most CEMP-s stars do accrete small amounts of material, consistent with the wind mass-transfer regime. 

\subsection{Mass distributions of post-accretion systems and their white-dwarf companions}

In Ba stars, an empirical relation similar to the binary mass function has been proposed to quantify correlations between the masses of the two system components: $$Q = \frac{M_{WD}^3}{(M_{Ba} + M_{WD})^2},$$ \noindent where $M_{Ba}$ is the mass of the barium star, and $M_{WD}$ is the mass of the white dwarf in solar masses. From \cite{2019A&A...626A.127J}, the value of $Q$ displays a bimodal distribution, with peaks at $Q = 0.057 \pm 0.009$ for the strong Ba stars and $Q = 0.036 \pm 0.027$ for the weak Ba stars. We tabulate our findings in Table \ref{tab:cap_Q}. In our sample of strong Ba stars, we find the median value of $Q = 0.055$. This distribution is highly non-Gaussian, with the mode of the strong Ba stars at $Q = 0.041$ and a second peak around $Q \approx 0.100$. For the weak Ba stars, we find a median value of $Q = 0.032$, where the mode of the distribution is at $Q = 0.025$. Both classes of Ba stars are within the bounds of the wide distribution determined by \citet{2019A&A...626A.127J}. 

We extend this to the metal-poor regime and investigate trends in the value of $Q$. In the combined sample of CEMP-s and CH stars, we find a median value of $Q = 0.043$, a prominent peak at a value of $Q = 0.037$, and a secondary peak at values $Q \approx 0.120$. As in the samples of Ba stars, the distribution in $Q$ is non-Gaussian. The secondary peaks in these populations could highlight the dwarf stars where the models are less constrained by the lack of mixing and dilution of surface material. 

\begin{table}[!h]
    \centering
    \caption{Computed Q values for the different populations of stars in our modeling efforts. The mode and mean averages are substantially different in the populations.}
    \begin{tabular}{c c c}
        \hline
        \hline
        Class & $\rm{median}(Q)$ & $\rm{mode}(Q)$ \\
        \hline
        Ba-weak     & 0.032 & 0.025 \\
        Ba-strong   & 0.055 & 0.041 \\
        CEMP-s + CH & 0.043 & 0.037 \\
        \hline
        \hline
    \end{tabular}
    \label{tab:cap_Q}
\end{table}

\subsection{Orbital properties}

\begin{figure}
    \centering
    \includegraphics[width=0.95\linewidth]{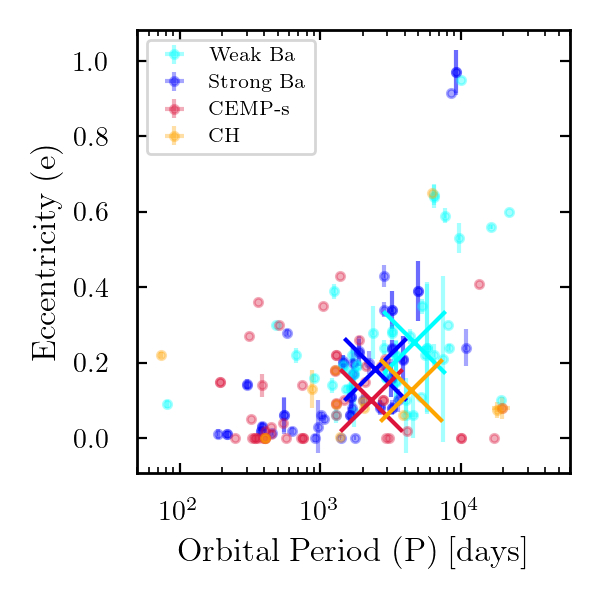}
    \caption{Eccentricity-period diagram for our combined sample of stars. Cyan data points are weak Ba stars, blue data points are strong Ba stars, orange data points are CH stars, and crimson data points are the CEMP-s stars. Centroids of the populations are marked with X's of corresponding colors.}
    \label{fig:ecc_per}
\end{figure}

Orbital periods and eccentricities, collected from \cite{1998A&A...332..877J},\citet{1998A&AS..131...25U}, \citet{2005ApJ...625..825L}, \citet{HansenT+_CEMP_binaries},\citet{Karinkuzi_Goswami2014}, \citet{Goswami_Aoki_Karinkuzhi2016},\citet{Jorissen+2016_BinaryCH_CEMP}, \citet{Goswami+2021_CEMPs}, and \citet{Dimoff+2024} are shown in Figure \ref{fig:ecc_per}, where cyan data points are weak Ba stars, blue data points are strong Ba stars, orange data points are CH stars, and red data points are the CEMP-s stars. There is significant scatter in the $e-P$ relation even within single populations, and many systems show high orbital eccentricities. We compute and mark the centroids of the populations of stars. The strong Ba stars and weak Ba stars trend toward slightly higher eccentricities ($\langle e \rangle=$0.25 and 0.18, respectively) compared to the CEMP-s and CH stars ($\langle e \rangle=$0.11 and 0.12, respectively). The weak Ba stars and CH stars have on average longer periods ($P \approx 4500$ days) compared to the strong Ba stars and the CEMP-s stars ($P \approx 2300$ days), although the spread in these data overlap with each other. Mass transfer should aid in circularizing the binary orbits through tidal interactions and angular momentum, and the older, more metal-poor populations show orbital eccentricities closer to zero than the younger, more metal-rich populations. \citet{2025arXiv250410939K} have made progress explaining the spread of orbital eccentricities in these post-accretion systems with the inclusion of a circumbinary disc to assist in angular momentum transport and enhancing the orbital eccentricity. 

When applying the accretion mass as a dimension to the $e-P$ plane, we find no significant trends with orbital period or eccentricity. We identify stars with robust abundance patterns in the large RV sample from \citet{Dimoff+2024} with long orbital periods ($P \geq 1000$ days) to model in deeper complexity using the \texttt{STARS} code in a follow-up paper; this list includes HD 201657, HD 211954, HD 224959, and HD 88562. 
\section{Conclusions}\label{sec:CONCLUSION}

In this study, we present a grid of binary accretion models for low-mass binaries. We find that the most chosen model to explain weak Ba stars is system that results in a $2.50$ \msun $\,$companion, accreting a moderate amount of mass $m_{acc} \lesssim 0.40$ \msun $\,$from a $2.00 - 3.00$ \msun $\,$AGB star. Strong Ba stars tend have lower initial masses $\sim 0.6$ \msun$\,$and accrete more material $m_{acc} \geq 0.50$ \msun $\,$from $2.50$ \msun$\,$AGB stars, resulting in higher accretion efficiencies at lower initial mass ratios. Both classes of Ba stars accrete moderate to high amounts of AGB material to explain the observed surface abundance patterns. This could be due to the higher metallicities, or higher final mass where in a more massive star the convective envelope mixes more strongly than in low mass stars. In both cases, wind mass-transfer models cannot explain the high accretion masses and efficiencies. Further modeling considering higher accretion masses may reveal a scenario that better describes the final mass distribution of the strong Ba stars. 

The CH and CEMP stars are similar in many ways. They share a preference for a $2.00 - 3.00$ \msun$\,$AGB stars and both populations show bimodal behavior with peaks at lower ($\sim0.80$ / $\sim1.00$ \msun) and higher masses ($\sim2$ / $\sim3$ \msun) in initial and final mass respectively. The initial mass ratios for these metal-poor and carbon-rich systems are also somewhat bimodally split between $q \approx 0.2$ and $q = 1$, and they tend to accrete less material ($\sim 0.1$ \msun$\,$) from their former AGB companion compared to the metal-rich systems. These stars display lower accretion efficiencies in better agreement with the wind regime at higher mass ratios, supported by their long orbital periods and comparatively lower eccentricities. Evolution modeling including rotation is planned to track the angular momentum through the accretion process. 

\begin{acknowledgements}
      This project has received funding from the European Union’s Horizon 2020 research and innovation programme under grant agreement No 101008324 (ChETEC-INFRA). We graciously thank ChETEC-INFRA TNA program for providing the framework and infrastructure. 
      Further support in funding comes from the State of Hesse within the Research Cluster ELEMENTS (Project ID 541 500/10.006) and HFHF, the Helmholz Research Academy Hessen for FAIR. 
      We acknowledge the support of the Data Science group at the Max Planck Institute for Astronomy (MPIA) and especially Dr. Raphael Hviding and Dr. Iva Momcheva for their invaluable assistance in developing the software for this research paper. We thank our colleague Dr. Sophie van Eck for their assistance in focusing the scope of this investigation and providing feedback on this manuscript.
\end{acknowledgements}

\bibliographystyle{aa}
\bibliography{references}{}

\begin{appendix} 

\section{The Model Grid}\label{apx:the_grid}

\noindent\begin{minipage}{\textwidth}
    \centering
    \includegraphics[width=0.98\linewidth]{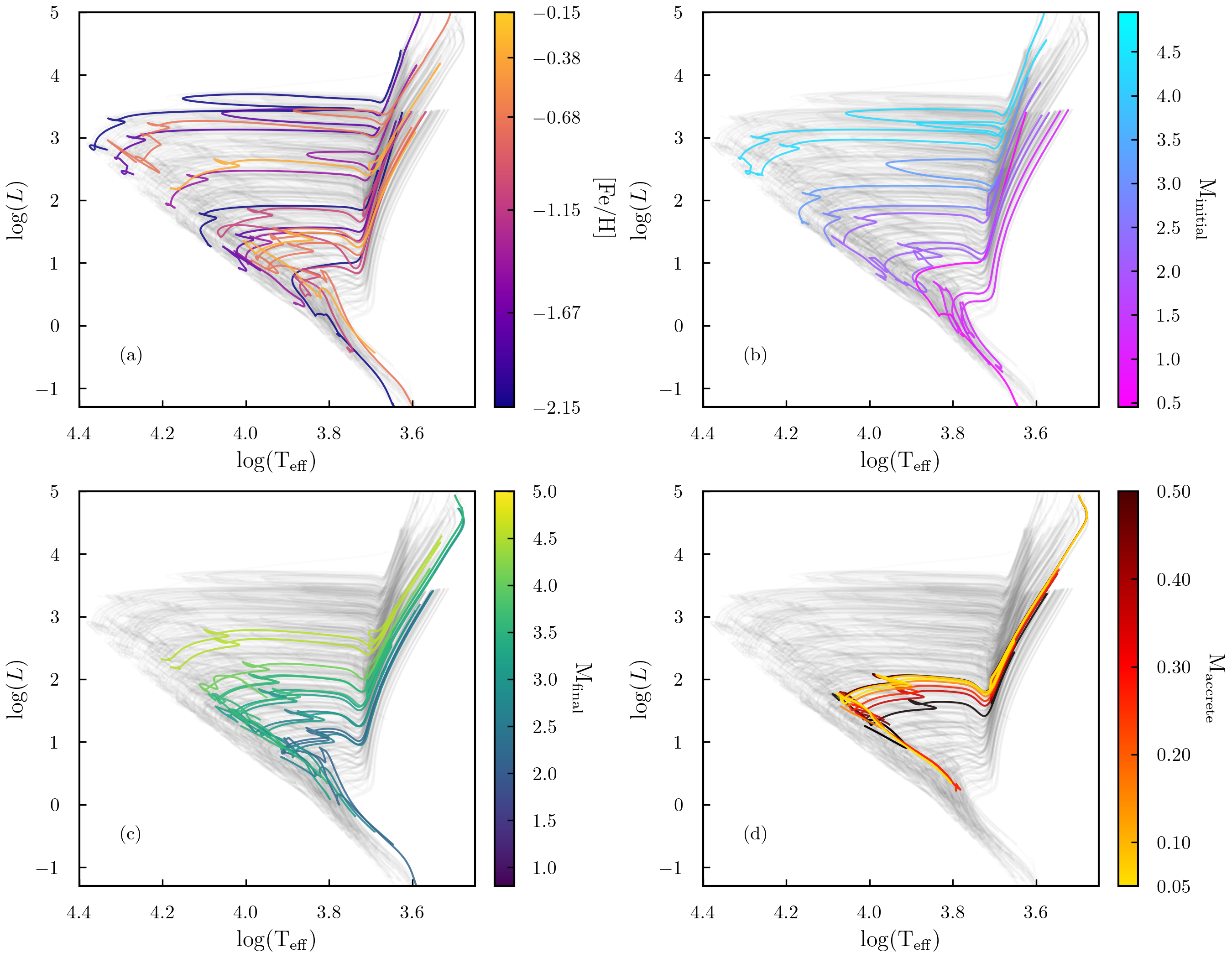}
    \captionof{figure}{Our computed grid of evolutionary models in the HR diagram, displaying every 5th model in gray. Selected models are highlighted to show the range of parameters across the grid. Panel (a) shows the varying metallicities of our models, with purple models at lower metallicity and orange models at higher metallicities. These are spread across the Kiel diagram. On the giant branch, metal-rich evolutionary tracks fall to the right side of the giant branch towards cooler temperatures, where metal-poor giants are on the left side at higher temperatures. Panel (b) shows the range of initial masses, with lower initial masses in blue and higher initial masses in pink. Lower initial-mass models appear towards the bottom and right of the panel, and higher initial-mass models to top and the left. Panel (c) shows the range of final masses. The highlighted evolutionary tracks are for a fixed metallicity of -0.15, with green and yellow models showing higher mass and blue models showing lower mass. This parameter follows the same general trend as the initial masses. Panel (d) displays an example of the varying accretion mass in our models. The highlighted tracks begin at different initial masses all resulting in $m_{final} = 2.50$ \msun, at a metallicity of [Fe/H] = -0.15. Black tracks show higher accretion masses, and yellow tracks show lower amounts of accretion.}\label{fig:the_grid}
\end{minipage}

\end{appendix}

\end{document}